\documentclass[10pt]{article}
\usepackage{fa2025}
\usepackage{amsmath}
\usepackage{cite}
\usepackage{url}
\usepackage{graphicx}
\usepackage{color}
\usepackage{siunitx}
\usepackage[utf8]{inputenc}
\usepackage{empheq}
\usepackage{amssymb}

\newcommand{\bA}{{\mathbf{A}}}

\newcommand{\bw}{{\mathbf{w}}}

\newcommand{\bx}{{\mathbf{x}}}
\newcommand{\bh}{{\mathbf{h}}}
\newcommand{\bP}{{\mathbf{P}}}
\newcommand{\bQ}{{\mathbf{Q}}}

\newcommand{\bk}{{\mathbf{k}}}
\newcommand{\bI}{{\mathbf{I}}}

\newcommand{\bD}{{\mathbf{D}}}

\newcommand{\bW}{{\mathbf{W}}}

\DeclareMathOperator{\sinc}{sinc}

\title{Tracking of Spatially Dynamic Room Impulse Responses Along Locally Linearized Trajectories}

\multauthor
{Kathleen MacWilliam$^{1*}$ \hspace{1cm} Thomas Dietzen$^{2}$ \hspace{1cm} Toon van Waterschoot$^1$} { \bfseries{}\\ $^1$ Department of Electrical Engineering (ESAT-STADIUS), KU Leuven, Belgium \\$^2$ Department of Electrical Engineering (ESAT-PSI), EAVISE, KU Leuven, Belgium\\ \correspondingauthor{kathleen.macwilliam@esat.kuleuven.be}{First author et al.}}

\begin{document}

\maketitle
\begin{abstract}
Measuring room impulse responses (RIRs) at multiple spatial points is a time-consuming task, while simulations require detailed knowledge of the room’s acoustic environment. In prior work, we proposed a method for estimating the early part of RIRs along a linear trajectory in a time-varying acoustic scenario involving a static sound source and a microphone moving at constant velocity. This approach relies on measured RIRs at the start and end points of the trajectory and assumes that the time intervals occupied by the direct sound and individual reflections along the trajectory are non-overlapping. The method’s applicability is therefore restricted to relatively small areas within a room, and its performance has yet to be validated with real-world data. 
In this paper, we propose a practical extension of the method to more realistic scenarios by segmenting longer trajectories into smaller linear intervals where the assumptions approximately hold. Applying the method piecewise along these segments extends its applicability to more complex room environments. We demonstrate its effectiveness using the trajectoRIR database, which includes moving microphone recordings and RIR measurements at discrete points along a controlled L-shaped trajectory in a real room.
\end{abstract}

\section{Introduction}
Knowledge of room impulse responses (RIRs) is an important aspect of room acoustics, with applications ranging from spatial audio rendering to speech enhancement and dereverberation \cite{Ratnarajah2022TowardsIR, borra2019soundfield}. However, measuring RIRs at multiple spatial positions is time-consuming, and while computational simulations offer an alternative, they typically require detailed knowledge of the room’s geometry and material properties, which is often unavailable or difficult to obtain. A practical alternative is to estimate RIRs along a trajectory using a moving microphone, allowing for efficient spatial sampling of the acoustic field. 

In our previous work \cite{macwilliam2024state}, we introduced a method for estimating the early part of RIRs along a linear trajectory in a time-varying acoustic scenario, involving a static source and a moving microphone. This can be framed as a time-varying system identification problem, where the system to be identified is referred to as a time-variant RIR. Here, we define a time-variant RIR as the linear time-invariant (LTI) system relating the source signal to the microphone signal if the microphone location is time-variant. 

Time-variant RIRs have been estimated using fully data-driven methods, such as normalized least mean squares (NLMS) algorithms \cite{Enzner2008, Antweiler2012}, and Kalman filtering approaches \cite{enzner2010bayesian}. In contrast, methods that operate with limited data often incorporate prior knowledge through an underlying room acoustic model. If the time-variant RIR is considered as a sequence of RIRs at discrete spatial positions, the estimation problem can be viewed as one of interpolation, a topic that has been explored in numerous studies \cite{Hahmann2022ACP, Karakonstantis2024RoomIR, antonello2017room, sundstrom2024optimal}. 
Most relevant to our scenario is the task of interpolating an RIR at a location between two measured positions, given RIRs from a common static source. This has been addressed using techniques such as Dynamic Time Warping (DTW), which temporally aligns early reflections prior to linear interpolation \cite{kearney2009dynamic, bruschi2020innovative}, and optimal transport-based methods, which allow non-bijective matching of reflection components \cite{geldert2023interpolation}.

In our prior work, we integrated RIR interpolation, derived from a room acoustic model, into a state-space framework. This approach combines physical modeling with data-driven estimation by using a first-order difference equation to model the evolution of early reflections over time. Unlike other approaches that use a state transition factor \cite{enzner2010bayesian, Nophut2024}, we incorporated interpolation into the state transition matrix between consecutive early RIR segments, derived from an analytical model for this transition based on the image source model (ISM) \cite{allen1979image}.
Along with a linear trajectory and constant velocity, our method assumes that the time intervals occupied by the direct sound and individual reflections over the entire trajectory do not overlap. These assumptions limit the method's applicability to small regions within a room, and its effectiveness in real-world conditions remains to be validated. Furthermore, the method relies on known RIRs at the start and end of the trajectory, from which the times of arrival (TOAs) of corresponding reflections must be accurately extracted and paired. To facilitate this, we proposed a DTW-based approach for estimating the parameters required to construct the derived transition matrix from the endpoint RIRs 

In this paper, we extend our earlier work by introducing a segmentation-based approach to accommodate more complex microphone trajectories. Specifically, we divide longer trajectories into smaller linear segments where the original assumptions approximately hold, allowing for a piecewise application of our method. This segmentation strategy enables the handling of non-constant microphone velocities and curved paths while maintaining the model’s validity. We validate both the original and extended methods using the trajectoRIR dataset \cite{trajectoRIR}, which includes moving microphone recordings along an L-shaped path in a real room, along with measured RIRs at discrete locations. We also emphasize the advantages of DTW-based transition matrix estimation in real-world scenarios, where inconsistencies in reflection amplitudes and the absence of certain reflections across RIRs are common.

\section{Signal Model}
Consider a microphone moving along a trajectory and recording a signal from a static source. Let \( k \) be the discrete-time index, and let \( l \) be a one-dimensional index representing the microphone's location along the trajectory. The RIR is the LTI system relating the source signal \( x(k) \) to the observed signal \( y(l, k) \) at location $l$. In this paper, we focus only on the early part of the RIR, excluding the reverberant tail, though we will still refer to them as RIRs throughout for simplicity.  
Let \( h(l, n) \) denote the early RIR at microphone location \( l \), where \( n \) indexes the time shift of the RIR samples. Defining the vectors:
\[
\bx(k) = \begin{pmatrix} x(k) & x(k-1) & \dots & x(k - N + 1) \end{pmatrix}^T,
\]
\[
\bh(l) = \begin{pmatrix} h(l,0) & h(l,1) & \dots & h(l,N-1) \end{pmatrix}^T.
\]
The observed signal is then modeled as  
\begin{align}
 y(l,k) &=\bx^T(k)\bh(l) + v(k),
   \label{definitionRIR_vec}
\end{align}
where \( v(k) \) represents noise and may also account for the late reverberation excluded from the RIR model.
If the microphone moves along a linear trajectory at a constant speed, the location index $l$ can be directly related to the time index $k$, as detailed in the next section.  For a nonlinear trajectory, we can approximate it as a sequence of locally linear segments, where each segment adheres to the same model assumptions. 
In this paper, we aim to estimate \( \bh(l) \) using the known source signal \( x(k) \) and the observed signal \( y(l, k) \), while accounting for a time-variant microphone location.

\section{Proposed State-Space Model}\label{PSS}
\begin{figure*}[ht!]
 \centerline{
 \includegraphics[width=17.0cm]{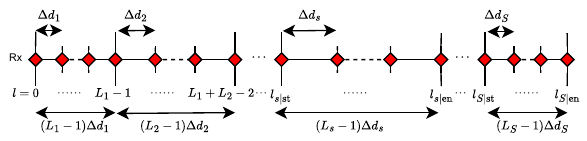}}
 \caption{Illustration of trajectory partitioned into \( S \) approximately linear segments of varying lengths.}
 \label{fig:trajmodel}
\end{figure*}

Now, consider a trajectory of length \( d \), which may be nonlinear, indexed by \( l \in \{0, \dots, L_{\text{total}} - 1\} \), where \( L_{\text{total}} \) is the total number of sampled locations. Let us partition the trajectory into \( S \) approximately linear segments of varying lengths, indexed by \( s \in \{1, \dots, S\} \). Within each segment, we assume the microphone moves with a constant velocity \( \mu_s \). Let segment \( s \) have length \( d_s \) and contain \( L_s \) uniformly spaced locations, where the first location of each segment coincides with the last location of the preceding segment. The spacing between consecutive locations within a segment is denoted by \( \Delta d_s \). The set of location indices for segment \( s \) is defined as  
\begin{equation}
    \mathcal{L}_s = \{ l_{s|\text{st}}, \dots, l_{s|\text{en}} \}, \quad s \in \{1, \dots, S\},
    \label{range}
\end{equation}  
where \( l_{s|\text{st}} \) and \( l_{s|\text{en}} \) are, respectively, the start and end location indices of the segment. Figure \ref{fig:trajmodel} illustrates this setup.  
%
Given a temporal sampling period \( T \), the smallest spatial increment within segment \( s \) for a constant velocity $\mu_s$ is $\Delta d_s = \mu_s T$. In this case, the microphone crosses one location \( l \) at each time instant \( k \), so the location-variant RIR \( \bh(l) \) can equivalently be interpreted as a time-variant RIR \( \bh(k) \) with \( k = l \). Accordingly, we define $y(l) = y(l,k)$, $\bx(l) = \bx(k)$ and $v(l) = v(k),$ allowing us to express the system in the following state-space form:  
\begin{align}
    \bh(l) &= \bA(l) \bh(l-1) + \bw(l),\label{state_equation_prop}\\
    y(l) &= \bx^T(l) \bh(l) + v(l),
   \label{observation_equation_prop}
\end{align}  
for \( l \in \{1, \dots, L_{\text{total}} - 1\} \).  
Here, \( \bA(l) \) is a location-variant transition matrix, and \( \bw(l) \) represents process noise modeling errors in \( \bA(l) \). Using this state-space model, an estimate \( \hat{\bh}(l) \) of \( {\bh}(l) \) can be obtained via a Kalman filter \cite{simon2006optimal}, with one recursion per location, as detailed in Section \ref{kalman}.  

To define \( \bA(l) \), we assume that within each approximately linear segment, the RIR at location \( l \) can be related to its predecessor through a segment-dependent transition matrix \( \bA_s \), which is location-\textit{invariant} within the segment i.e.,  
\begin{equation}
    \bh(l) \approx \bA_s \bh(l-1), \quad l \in \mathcal{L}_s \setminus \{l_{s|\text{st}}\}.
    \label{tranConst}
\end{equation}  
We thus propose that the location-variant transition matrix $\bA(l)$ be defined as a piecewise constant function that follows the segment structure, such that
\begin{equation}
\bA(l) =
\begin{cases} 
\bA_1, & l \in \mathcal{L}_1 \setminus \{l_{1|\text{st}}\} \\
\bA_2, & l \in \mathcal{L}_2 \setminus \{l_{2|\text{st}}\}\\
\vdots \\
\bA_S, & l \in \mathcal{L}_S \setminus \{l_{S|\text{st}}\}
\end{cases}
\label{segmentedA}
\end{equation}   
The segment-dependent matrix \( \bA_s \) can be derived using the ISM and estimated by interpolating the TOAs of the direct component and reflections between assumed known RIRs at the segment boundaries. This is detailed in Section \ref{DTW_matrix}. 


\subsection{Kalman filter update equations}\label{kalman}
For the state-space model in \eqref{state_equation_prop}–\eqref{observation_equation_prop}, let
$\bQ = \operatorname{E}[\bw(l)\bw^T(l)]$ and $R = \operatorname{E}[v^2(l)]$ denote the covariance of the process noise \( \bw(l) \) and the variance of the observation noise \( v(l) \), respectively. While \( R \) can potentially be estimated from background noise recordings, \( \bQ \) is typically a tuning parameter. The Kalman filter update equations are given by:  
\begin{align}
    \hat{\bh}(l) & = \bA(l)\hat{\bh}^{+}(l-1), \label{eq:state-prediction} \\
    \bP(l) & = \bA(l)\bP^{+}(l-1) \bA^T + \bQ, \label{eq:covariance-prediction} \\[5pt]
    \bk(l) & = \dfrac{\bP(l) \bx(l)}{\bx^T(l) \bP(l) \bx(l) + R}, \label{eq:kalman-gain} \\
    \hat{\bh}^{+}(l) & = \hat{\bh}(l) + \bk(l) \left(y(l) - \bx^T(l) \hat{\bh}(l)\right), \label{eq:state-update} \\
    \bP^{+}(l) & = \left(\bI - \bk(l) \bx^T(l)\right) \bP(l), \label{eq:covariance-update}
\end{align}
Equations \eqref{eq:state-prediction}–\eqref{eq:covariance-prediction} compute the prior estimates (prediction step), while \eqref{eq:kalman-gain}–\eqref{eq:covariance-update} refine them into posterior estimates, denoted by the superscript $^+$, using new observations (update step). Here \( \hat{\bh}(l) \) and \( \hat{\bh}^{+}(l) \) are the state estimates, \( \bP(l) \) and \( \bP^{+}(l) \) are the respective state-estimation error covariance matrix estimates, and \( \bk(l) \) is the Kalman gain. The filter is initialized with \( \hat{\bh}^{+}(0) = \bh(0) \), assuming a known initial RIR, and \( \bP^{+}(0) \) is set to a small-norm matrix, reflecting negligible initial uncertainty.  
Unlike common state-space models, such as the one in \cite{enzner2010bayesian}, which use a scalar $\alpha$ instead of \( \bA(l) \), we estimate the segment-dependent transition matrix \( \bA_s \) needed to construct $\bA(l)$ as in \eqref{segmentedA} using the ISM-based approach detailed in the next section. If no recorded signals are available, i.e., \( y(l) = 0 \) and \( \bx(l) = 0 \), the recursion simplifies to 
\begin{equation}
\hat{\bh}(l) = \bA(l)\hat{\bh}(l-1) \approx  \bA_s^{(l-l_{s|\text{st}})} \bh(l_{s|\text{st}}),
\label{interop}
\end{equation} 
which corresponds to a linear interpolation of the RIRs, similar to the methods in \cite{kearney2009dynamic, geldert2023interpolation}.

\subsection{Segment-dependent transition matrix}\label{DTW_matrix}

Here we present a summary of the ISM-based model proposed in \cite{macwilliam2024state} for the transition matrix $\bA_s \in  \mathbb{R}^{N \times N}$ that approximates the mapping between two RIRs $\bh(l_{s|\text{st}})$ and $\bh(l_{s|\text{en}})$ as in equation \eqref{tranConst}.
Using the ISM, the RIR can be expressed as the sum of contributions from the original sound source and additional image sources, which represent reflections within the room. 
Let the ISM representation of the discrete-time RIR  at location $l$ with discrete time-shift index $n$ and sampling period \(T\) be given by,
\begin{equation}
h(l, n) = \sum_{r \in \mathcal{R} } a_{r}(l) \sinc \left (n-\frac{\tau_{r}(l)}{T}\right)
\label{ISMdiscrete}
\end{equation}
where $r \in \mathcal{R}=\{0,1,..R-1\}$ is the (image) source or reflection index, and $\tau_{r}(l)$ and $a_{r}(l)$ respectively denote the TOA and amplitude of the (image) source $r$ observed at location $l$. 
Let 
\begin{align}
\Delta_{r}(l) =\tau_{r}(l)-\tau_{r}(l-1)
\label{TDOA}
\end{align}
be the time difference of arrival (TDOA) of reflection $r$ between locations $l$ and $l-1$.
Recall that we are considering a linear segment of the trajectory with $l \in \mathcal{L}_s \setminus \{l_{s|\text{st}}\}$ where $\mathcal{L}_s$ is defined in \eqref{range}. Under the assumption of far-field conditions, where the (image) sources are located much further away than the length of the segment, we get ${a_{r}(l)}={a_{r}(l-1)}$ and ${\Delta}_r(l) = {\Delta}_{r,s}$. We can obtain an estimate of this constant TOA shift using 
\begin{equation}
{\Delta}_{r,s} = \frac{\tau_r(l_{s|\text{en}}) - \tau_r(l_{s|\text{st}})}{L_s-1},
\label{estimatedTDOAs}
\end{equation}
where \( \tau_r(l_{s|\text{st}}) \) and \( \tau_r(l_{s|\text{en}}) \) denote the (image) source TOAs at the initial and final locations on the segment respectively, and \( L_s \) is the total number of locations on that segment.
Let us now define the interval that the TOAs of non-negligible components of reflection $r$ occupy over the segment $l \in \mathcal{L}_s \setminus \{l_{s|\text{st}}\}$ as
\begin{align}
 \mathcal{T}_{r,s} &= [{\tau}_{r,s|\operatorname{min}},\, {\tau}_{r,s|\operatorname{max}}]  \\
 {\tau}_{r,s|\operatorname{min}} &= \operatorname{min}\bigl(\tau_r(l_{s|\text{st}}+1),\, \tau_r(l_{s|\text{en}})  \bigr) - \epsilon,  \\
 {\tau}_{r,s|\operatorname{max}} &= \operatorname{max}\bigl(\tau_r(l_{s|\text{st}}+1),\, \tau_r(l_{s|\text{en}})  \bigr) + \epsilon.
 \label{interval_Tr}
\end{align}
where $2\epsilon$ is the effective temporal width of an individual reflection $r$. Based on this we define the set
\begin{equation}
\boxed{\tilde{\mathcal{R}}_s(n) = \left\{r \in \mathcal{R}\,\left|\, n \in \frac{\mathcal{T}_{r,s}}{T} \right.\right\}} .
 \label{subset2}
\end{equation}
Under the further assumption that the intervals occupied by individual reflections along the segment do not overlap, we derived the relation between \( h(l,n) \) and \( h(l-1, n') \) in \cite{macwilliam2024state}. Here, \( n \) and \( n' \) denote the respective time shift indices, and the relation is given by
\begin{equation}
 h(l,n) \approx \sum_{n'} \sum_{r\in \tilde{\mathcal{R}}_s(n)}  \sinc \left( n- \frac{\Delta_{r,s}}{T}-n' \right)  h(l-1, n')
 \label{h2h1_inv}
\end{equation}
for $l \in \mathcal{L}_s \setminus \{l_{s|\text{st}}\}$. Consequently, the element of $\bA_s$ at index $(n+1, n'+1)$ is:
\begin{equation}
\boxed{A_{s(n+1,n'+1)}= \sum_{r\in \tilde{\mathcal{R}}_s(n)} \sinc \left( n- \frac{\Delta_{r,s}}{T}-n' \right).}
\label{elements_transition_matrix_invariant}
\end{equation}
when $\tilde{\mathcal{R}}_s(n) \neq \varnothing$, and $A_{s( {n+1,n'+1})} = 0$ otherwise. A detailed derivation is available in \cite{macwilliam2024state}.
\subsection{Practical estimation of transition matrix }\label{dtw_section}
\begin{figure}[t]
 \centerline{
 \includegraphics[width=7.8cm]{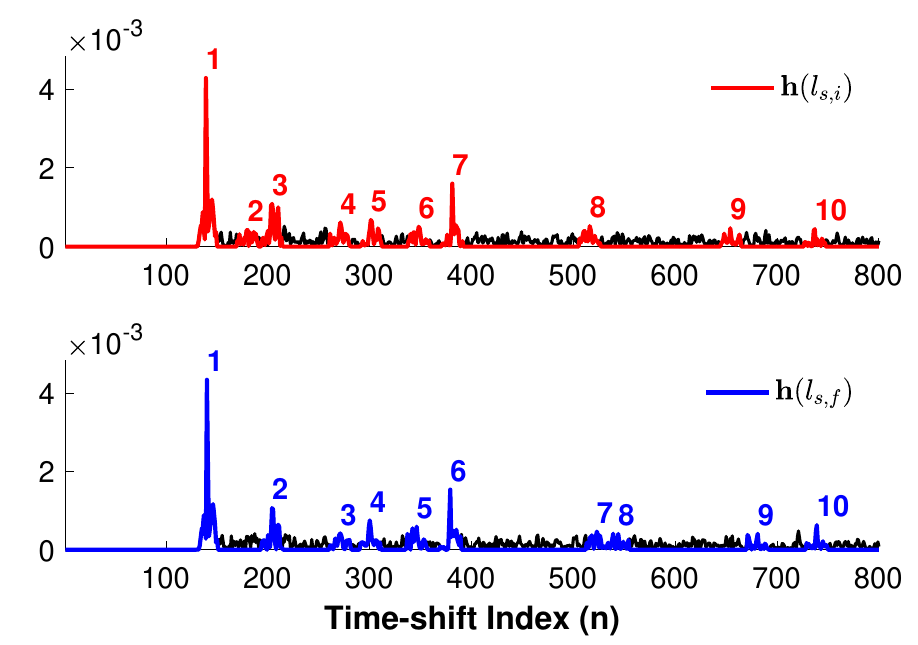}}
 \caption{Example of mismatched TOA estimations using peak-picking and a matched filter approach as in \cite{6255766,6809961}.}
 \label{fig:mismatchedTOAs}
\end{figure}

The proposed transition matrix in \eqref{elements_transition_matrix_invariant} relies on accurately retrieving and pairing TOAs in \(\bh(l_{s|\text{st}})\) and \(\bh(l_{s|\text{en}})\). Assuming non-overlapping TOA intervals, reflections should arrive in the same order. A common method for identifying TOAs is to select the $R$ largest peaks using a peak-picking algorithm \cite{brookes1997voicebox} after matched filtering \cite{6255766}, refined via DTW in \cite{6809961} to improve peak detection of reflections that poorly correlate with the direct sound due to shape changes. However, owing to variations in reflection amplitudes or the absence of certain reflections at specific locations (particularly in real data), the selected peaks may differ between RIRs. Consequently, reflections may be incorrectly paired, as illustrated in Figure \ref{fig:mismatchedTOAs}.
Therefore, a method was proposed to directly estimate the parameters needed for the matrix \(\mathbf{A}_s\) using DTW. This approach aligns elements in \(\bh(l_{s|\text{st}})\) and \(\bh(l_{s|\text{en}})\), leveraging their warp matrices to extract both \(\tilde{\mathcal{R}}_s(n)\) and \(\tau_r(l_{s|\text{en}}) - \tau_r(l_{s|\text{st}})\), the latter of which can be used to estimate \(\Delta_{r,s}\) via \eqref{estimatedTDOAs}.

The DTW algorithm \cite{muller2007dynamic} computes an accumulated cost matrix \(\bD \in \mathbb{R}^{(N+1) \times (N+1)}\), where each entry represents the cost of aligning \(h(l_{s|\text{en}},n)\) with \(h(l_{s|\text{st}},n')\). The cost function is defined as:
\begin{align*}
\resizebox{\columnwidth}{!}{$
D_{n+2, n'+2} = \lVert h(l_{s|\text{en}},n)-h(l_{s|\text{st}},n')\rVert_2 
+ \min \left\{ \begin{array}{c}
D_{n+1, n'+2} \\ 
D_{n+2, n'+1} \\ 
D_{n+1, n'+1}
\end{array} \right\}
$} 
\end{align*}
initialized \footnote{An additional row and column are used exclusively for initialization. As matrices in this paper are indexed starting from (1,1) while RIR sequences are indexed from 0, the entries of the accumulated cost matrix are therefore indexed as \( D_{n+2, n'+2} \).} with \(D_{1,1}=0\) and boundary values set to \(\infty\).
The optimal alignment is represented by a warp path $w= \left\{ (n+2, n'+2)\right\}_{i=1}^{I}$, a sequence of $I$ index pairs that defines correspondences between the samples in the RIR vectors. This path is found by backtracking from \(D_{N+2, N+2}\) to \(D_{2,2}\), maintaining a monotonically increasing order.  Note that a single point in one RIR can correspond to multiple points in the other, thus $I \ge N$. 
Once the optimal alignment is determined, we can warp the RIRs to align them temporally. Specifically, \(\bh(l_{s|\text{st}})\) can be warped to align with \(\bh(l_{s|\text{en}})\), yielding \(\tilde{\bh}(l_{s|\text{st}})\), and vice versa, yielding \(\tilde{\bh}(l_{s|\text{en}})\). This warping process can be expressed as a matrix operation on each of the original vectors namely,
\begin{align*}
\tilde{\bh}(l_{s|\text{st}}) = \bW_{\text{st}} \bh(l_{s|\text{st}}), \quad
\tilde{\bh}(l_{s|\text{en}}) = \bW_{\text{en}} \bh(l_{s|\text{en}}),
\end{align*}
where \(\bW_{\text{st}}, \bW_{\text{en}} \in \mathbb{R}^{I \times N}\) are sparse matrices constructed such that \([\bW_{\text{st}}]_{i, n+2} = 1\) and \([\bW_{\text{en}}]_{i, n'+2} = 1\) for each \(i \in \{1, \dots, I\}\), with \((n+2, n'+2)\) being the \(i^\text{th}\) index pair in the warp path $w$; all other entries are zero. 
%
%
Now, in order to find a direct mapping from \( \mathbf{h}(l_{s|\text{st}}) \) to \( \mathbf{h}(l_{s|\text{en}}) \), we aim to satisfy the condition $\bW_{\text{st}} \bh(l_{s|\text{st}})\approx \bW_{\text{en}} \bh(l_{s|\text{en}})$. This problem can be formulated as a least squares problem, leading to 
\begin{align}
\bh(l_{s|\text{en}}) = \bW_{\text{en}}^{+} \bW_{\text{st}} \bh(l_{s|\text{st}})
\label{warp}
\end{align}
where $\bW_{\text{en}}^{+}$ is the generalized left inverse of $\bW_{\text{en}}$. For a more detailed derivation of \eqref{warp} we refer the interested reader to \cite{6809961}. The warp matrix product $\bW_{\text{en}}^{+} \bW_{\text{st}}$, which we hereby denote as $\bW$, serves as an estimate of \( \bA_s^{(l_{s|\text{en}}-l_{s|\text{st}})} \) for \( l = l_{s|\text{en}} \) in Equation \eqref{interop}
i.e.,
\begin{equation}
\bW=\bA_s^{(l_{s|\text{en}}-l_{s|\text{st}})} = \bA_s^{L_s -1}.
\end{equation} 
Finding an estimate of \(\bA_s\) by directly computing the \((L_s -1)^{\text{th}}\) root of \(\bW\) is not straightforward due to potential non-diagonalizability and numerical instability, therefore we instead analyze its structural properties. When a single point in \( \bh(l_{s|\text{st}}) \) maps to a single point in \( \bh(l_{s|\text{en}}) \), the corresponding row or column in $\bW$ contains only one nonzero entry. However, if multiple points align, multiple nonzero entries appear in that row or column. Therefore, diagonal segments in \( \bW \) typically indicate regions where similar patterns, such as reflections, align. 

\begin{figure}[t]
 \centerline{
 \includegraphics[width=7.8cm]{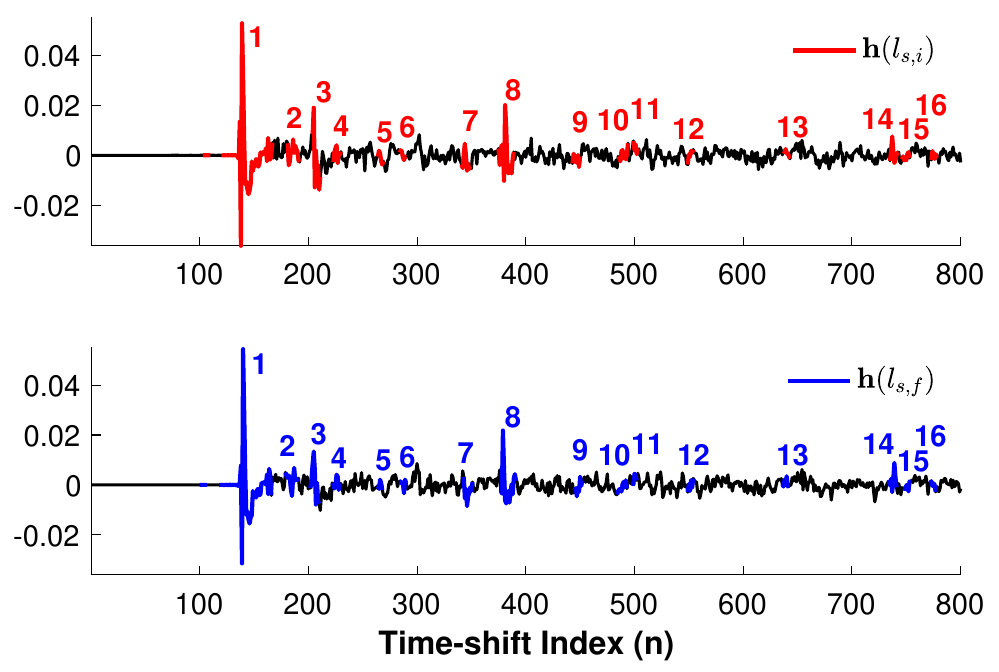}}
 \caption{Illustration of correctly paired peaks in RIRs using the proposed DTW-based method.}
 \label{fig:dtwTOAS}
\end{figure}

Using this observation, integer estimates of $(\tau_r(l_{s|\text{en}})-\tau_r(l_{s|\text{st}}))/T$ can be obtained by measuring the `distances' (number of elements) from sufficiently long diagonal segments in $\bW$ to the main diagonal. These estimates are then used to find estimates of $\Delta_{r,s}$ using the relation in \eqref{estimatedTDOAs} and the known sampling period $T$ leading to:
\begin{equation}
{\hat{\Delta}}_{r,s} = \frac{{\hat{\tau}}_r(l_{s|\text{en}}) - {\hat{\tau}}_r(l_{s|\text{st}})}{L_s-1}.
\label{eqyref}
\end{equation}
The start and end indices of the diagonal segments are also retrieved, denoted respectively as $(n_{r|\text{st}}, n_{r|\text{st}}')$ and $(n_{r|\text{en}}, n_{r|\text{en}}')$. These indices are then used to find a suitable estimate of the range $\mathcal{T}_{r,s}$ for the set $\tilde{\mathcal{R}}_s(n)$ in Equation \eqref{subset2} as,
\begin{align}
\hat{\mathcal{T}}_{r,s} &= [\hat{{\tau}}_{r,s|\operatorname{min}},\, \hat{{\tau}}_{r,s|\operatorname{max}}]  \label{interval_dtwTr1}\\
 \hat{{\tau}}_{r,s|\operatorname{min}} &= T\cdot \operatorname{min}\left(n'_{r|\text{st}} + \frac{\hat{\Delta}_{r,s}}{T},\, n_{r|\text{st}} \right),  \label{interval_dtwTr2}\\
\hat{{\tau}}_{r,s|\operatorname{max}} &= T \cdot \operatorname{max}\left(n_{r|\text{en}},\, n'_{r|\text{en}} + \frac{\hat{\Delta}_{r,s}}{T} \right).
 \label{interval_dtwTr3}
\end{align}
Figure \ref{fig:dtwTOAS} illustrates the correctly associated TOA ranges derived from the diagonal segments. Unlike the peak-picking method shown in Figure \ref{fig:mismatchedTOAs}, our DTW-based approach does not require specifying the number of peaks, as it automatically aligns all potential reflection peaks across the RIRs. A minimum diagonal segment length is enforced to reduce incorrect reflection peak detections.
Finally, we construct \(\hat{\bA}_s\) using Equations \eqref{elements_transition_matrix_invariant} and \eqref{subset2} with estimates $\hat{\Delta}_ {r,s}$ and $\hat{\mathcal{T}}_{r,s}$ obtained from Equations \eqref{eqyref} to \eqref{interval_dtwTr3}. Additionally, as a robustness measure for sections where reflections are not identified, ones are added along the main diagonal.

\section{Experimental evaluation and Results}
\begin{figure}[t]
 \centerline{
 \includegraphics[width=7.8cm]{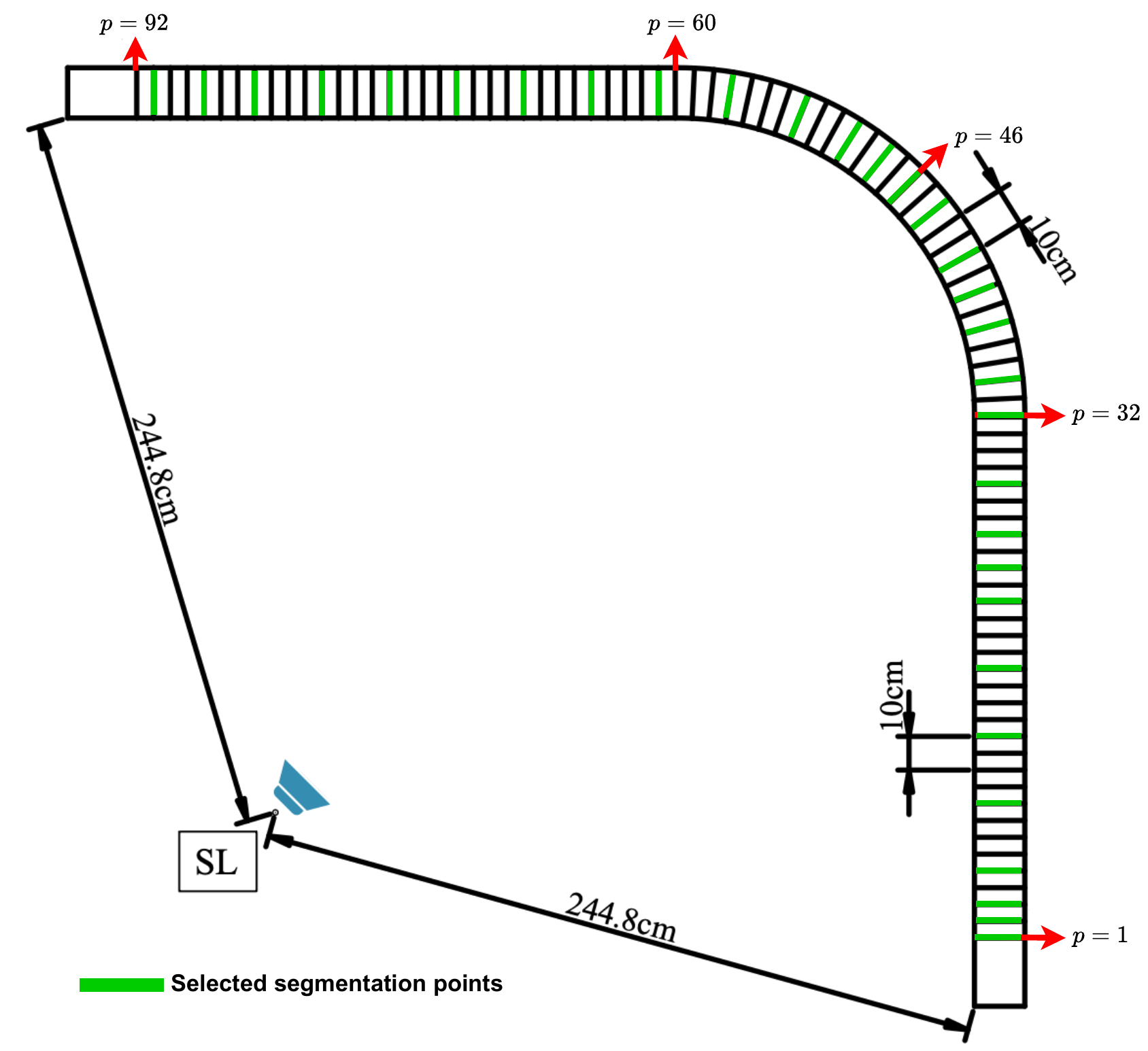}}
 \caption{Schematic of trajectory with points used for the segmentation in green.}
 \label{fig:traj}
\end{figure}

In this experiment, we evaluate the proposed method using real data from the trajectoRIR database \cite{trajectoRIR}. The setup we chose consists of a static sound source emitting a white noise signal while a microphone moves along an L-shaped trajectory at a constant velocity of \( 0.8 \) m/s, recording the signal. RIRs were also measured at fixed locations along the trajectory at 5 cm intervals, resulting in a total of 92 measurement points. We denote the set of all measurement indices as $\mathcal{P} = \{1, 2, \dots, 92\}$. A schematic of the trajectory with corresponding RIR measurement points is shown in Figure \ref{fig:traj}.
The initial recordings were made at a sampling rate of 48 kHz, but to align with typical speech processing applications, the signals were downsampled to 16 kHz. This downsampling also reduces the computational load as less data points need to be processed across the entire trajectory.
\subsection{Trajectory segmentation}
As the microphone moved through the fixed trajectory locations, the corresponding timestamps in the recorded signal, \( y(l) \), were retrieved as detailed in \cite{trajectoRIR}. We denote the corresponding location index of each measurement point \( p \in \mathcal{P} \) as \( l_p \). For segmentation purposes, we selected a subset of measurement points, \( \tilde{\mathcal{P}} \subseteq \mathcal{P} \), with $\tilde{\mathcal{P}} = \{ 1,2,3,5,9,13,17,21,23,25,28,32,34,37,39,41,44,\allowbreak 46,48,50,53,57,61,65,69,73,\allowbreak 77,81,85,88,91 \}$.

The selected measurement points (highlighted in green in Figure \ref{fig:traj}) are more densely spaced at the start and end of the trajectory, where velocity was not yet constant, and along the curved section. This ensures our modeling assumptions hold on those segments. Additionally, closer spacing was used in sections where the visibility of certain reflection peaks varied significantly if further-spaced RIRs were used (e.g., $p=[21,25]$). 
Based on the measured RIRs at the selected points, the segment-dependent transition matrices \( \mathbf{A}_s \) were then constructed as described in Section \ref{dtw_section}. 
%

%
\subsection{Kalman filter parameters}\label{kalman_parameters}
Given the assumption of knowledge of the RIR at the start of the trajectory $\mathbf{h}(0)$, the Kalman filter is initialized with $\hat{\mathbf{h}}^{+}(0) = \mathbf{h}(0)$. The measurement noise variance $R$ was not known in this case but after some initial testing of the Kalman filter was chosen as $R=0.01$, while the process noise covariance matrix $\mathbf{Q}$ was a tuning parameter chosen as $\mathbf{Q} = \sigma_w^2 \mathbf{I}$, where $\bI$ is an identity matrix and $10\log(\sigma_w^2) = -50$ dB.
The appropriate $\bA(l)$ matrix is selected for each algorithm, as outlined in Table \ref{tab:algorithms}. It was found that for the reference Kalman filter, a transition factor of $\alpha=1$ was suitable and varying this or applying a forgetting factor did not noticeably change the results.
\begin{table}[h]
    \caption{Description of the compared algorithms.}
    \centering
    \footnotesize 
    \begin{tabular}{|c|p{0.6\linewidth}|}
        \hline
        \rule{0pt}{13pt}\textbf{Algorithm Name} & \textbf{Algorithm Description} \\
        \hline
        \rule{0pt}{13pt} $\textbf{KF-}\alpha$ & Kalman Filter update equations \eqref{eq:state-prediction}-\eqref{eq:covariance-update} with $\bA(l)$ replaced with $\alpha=1$. \\[8pt]
        \hline
        \rule{0pt}{13pt} $\textbf{KF-}\bA(l)$ & Kalman Filter update equations \eqref{eq:state-prediction}-\eqref{eq:covariance-update} with $\bA(l)$ estimated using the segment-dependent $\hat{\bA}_s$  described in Section \ref{dtw_section}\\[8pt]
        \hline
        \rule{0pt}{13pt} $\textbf{LI-}\bA(l)$ & Linear interpolation equivalent to Kalman Filter update equation \eqref{eq:state-prediction} with $\bA(l)$ estimated using the segment-dependent $\hat{\bA}_s$ described in Section \ref{dtw_section} \\[8pt]
        \hline        
    \end{tabular}
    \label{tab:algorithms}
\end{table}

\subsection{Performance measures}
Since the measured RIRs are only available at discrete points along the trajectory, we evaluate the accuracy of the estimated location- or time-variant RIR at these specific locations. Let \(\hat{\mathbf{h}}^{+}(l_p)\) and \(\mathbf{h}(l_p)\) denote the estimated and measured RIRs at location index \(l_p\), respectively. To assess performance, we use a combination of two key metrics: normalized cross-correlation (NCC) for alignment and normalized misalignment (NM) for error quantification.  

To address small uncertainties in positioning during RIR measurement and tracking errors during microphone movement, NCC is initially used to correct for any resulting inaccuracies in selecting the appropriate location-variant RIR index corresponding to the measured RIR at \(l_p\). 
Specifically, we align \(\hat{\mathbf{h}}^{+}(l_p)\) with \(\mathbf{h}(l_p)\) by determining the delay \(\lambda_{\max}(l_p)\) that maximizes the NCC function,
\[
\lambda_{\max}(l_p) = \arg\max_{\lambda} \frac{ \mathbf{h}(l_p) \cdot \hat{\mathbf{h}}^{+}(l_p-\lambda)}{\|\mathbf{h}(l_p)\|_2 \|\hat{\mathbf{h}}^{+}(l_p-\lambda)\|_2}.
\]  
We denote the resulting aligned estimated RIR as $\hat{\mathbf{h}}^{+}(l_p - \lambda_{\max}(l_p))$ and then compute the NM in decibels to quantify the error:  
\begin{equation}
\mathcal{M}_{\operatorname{dB}}(l_p) = 20 \log_{10}\left(\frac{\lVert \hat{\mathbf{h}}^{+}(l_p - \lambda_{\max}(l_p)) - \mathbf{h}(l_p) \rVert_2}{\lVert \mathbf{h}(l_p) \rVert_2}\right).
\label{NM}
\end{equation}  
with lower values indicating a better match between the estimated and measured RIRs. 

Lastly, we evaluate how well the estimated location-variant RIR reproduces the recorded microphone signal. This is done by convolving the estimated location-variant RIR with the known input signal to obtain an estimated microphone signal, $\hat{y}(l) = \bx^T(l)\hat{\mathbf{h}}^{+}(l)$, and computing its correlation with the recorded microphone signal $y(l)$.

\subsection{Results and Discussion}
\begin{figure}[t]
 \centerline{
 \includegraphics[width=7.8cm]{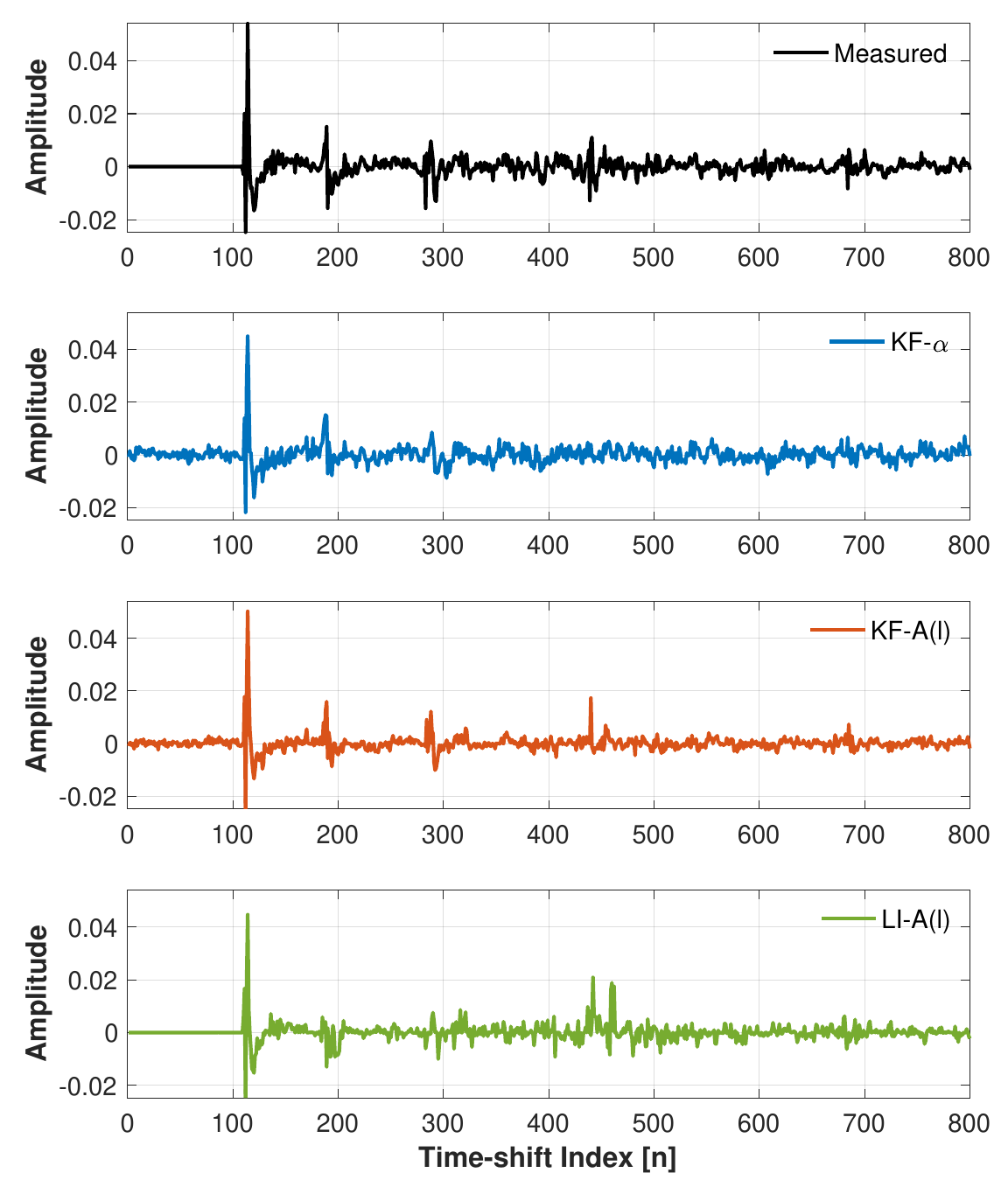}}
 \caption{Measured RIR at a point along the trajectory compared to estimated RIRs from various algorithms.}
 \label{fig:Results2}
\end{figure}

\begin{figure*}[t]
 \centerline{
 \includegraphics[width=17.0cm]{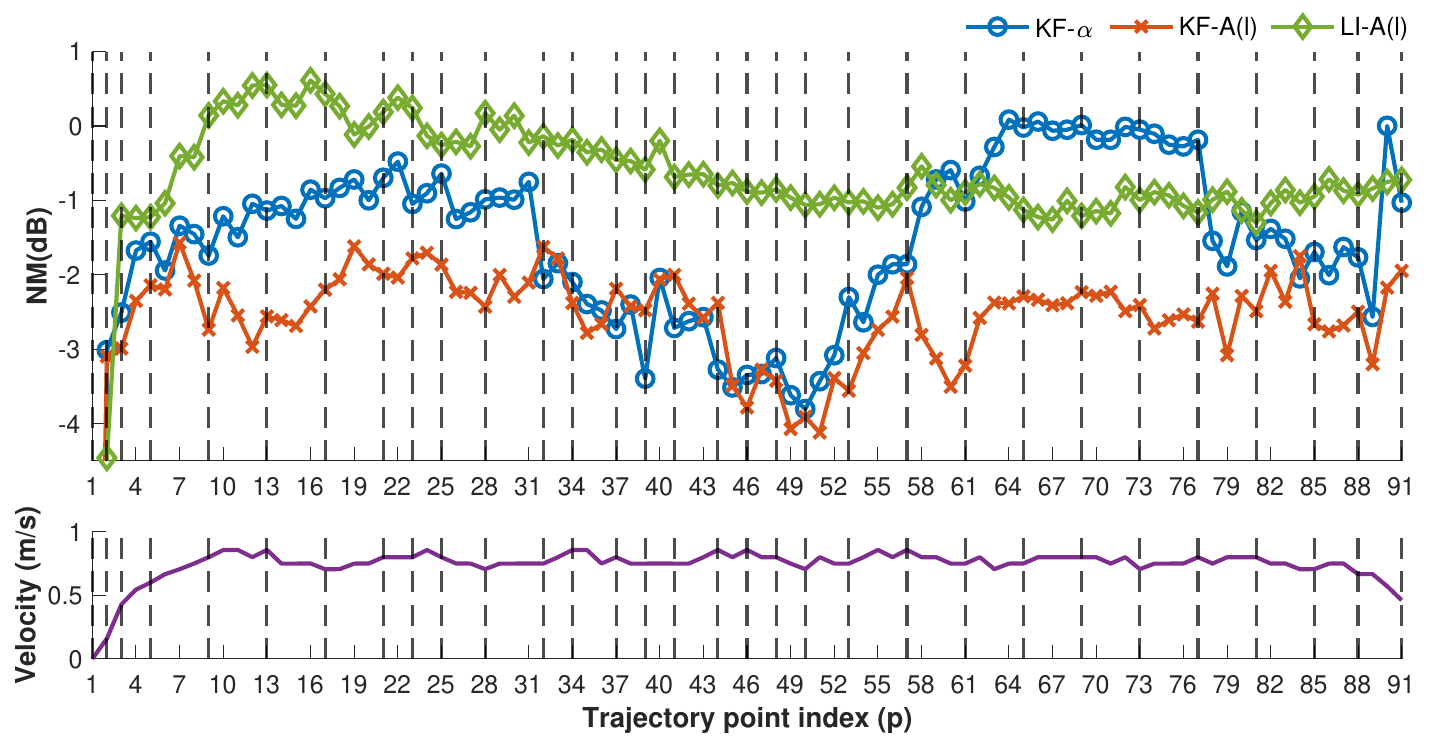}}
 \caption{Top: Normalized misalignment between estimated and measured RIRs along the trajectory. Bottom: Estimated average velocity between trajectory points.}
 \label{fig:Results1}
\end{figure*}

The top panel of Figure \ref{fig:Results1} illustrates the normalized misalignment of the different algorithms as defined in \eqref{NM}, with vertical dashed lines indicating segmentation points. The bottom panel shows the estimated average velocity of the microphone between each known trajectory point, derived from timestamp data and interpoint distances. Figure \ref{fig:Results2} shows an example of the measured RIR and corresponding estimated RIRs at a selected trajectory point, allowing a qualitative comparison of the different algorithms. Lastly, Table \ref{tab:correlation_results} contains the correlation coefficient between the estimated and actual recorded microphone signal for each algorithm.

\begin{table}
    \centering
    \small
    \caption{Correlation coefficients between estimated and recorded microphone signals for different algorithms.}
    \begin{tabular}{lc}
        \hline
        Algorithm & Correlation Coefficient \\
        \hline
        $\textbf{KF-}\alpha$ & 0.9382 \\
        $\textbf{KF-}\bA(l)$ & 0.9444 \\
        $\textbf{LI-}\bA(l)$ & 0.3359 \\
        \hline
    \end{tabular}
    \label{tab:correlation_results}
\end{table}

The results in Figure \ref{fig:Results1} indicate that the proposed algorithm, $\textbf{KF-}\bA(l)$, generally outperforms the reference methods $\textbf{LI-}\bA(l)$ and $\textbf{KF-}\alpha$. This performance advantage is particularly apparent in the linear sections of the trajectory, where velocity remains approximately constant, which is expected given these most strongly meet the original underlying model assumptions.

The $\textbf{LI-}\bA(l)$ algorithm consistently underperforms compared to both data-driven algorithms, underscoring the limitations of purely interpolative methods. However, between points $p = [60,78]$, both $\textbf{KF-}\alpha$ and $\textbf{KF-}\bA(l)$ see a degrade in performance, resulting in $\textbf{LI-}\bA(l)$ outperforming $\textbf{KF-}\alpha$. This decline stems from a spike in noise in the recorded microphone signal along this section of the trajectory, though the cause remains unknown. We retained the use of this particular signal to demonstrate our method’s robustness to noise variation, as it still achieves the best performance in these circumstances.

Around the midpoint of the trajectory (\(p=46\)), it can be observed that the performances of $\textbf{KF-}\alpha$ and $\textbf{KF-}\bA(l)$ somewhat converge. This can be attributed to the decrease in the rate of change in the TOA of the direct source as the microphone approaches the turning point \(p=46\) of its symmetrical curved path around the source position (see Figure \ref{fig:traj}). Under these conditions, the assumption of a transition factor of $\alpha=1$ becomes a reasonable approximation, explaining the similar performance between the two Kalman-based approaches.

Performance differences are also evident in the curved sections of the trajectory. In the range \(p=[46,60]\), $\textbf{KF-}\bA(l)$ outperforms the other methods, however, in the range \(p=[32,46]\), this is not always the case. This discrepancy may stem from significant variations in the amplitudes of associated reflection peaks in this region. A closer inspection of the measured RIRs confirmed this, revealing that certain reflections were either absent or significantly dampened at specific points along this section of the trajectory. This is likely related to the surrounding acoustic environment, particularly the narrow planar underside of the staircase above the trajectory \cite{trajectoRIR}.

While all methods perform comparably in estimating the direct path component, $\textbf{KF-}\bA(l)$ seems to have better accuracy in capturing reflection peaks, as evident in Figure \ref{fig:Results2}. This is particularly important as early reflections play a crucial role in intelligibility in speech applications, localization, and room geometry estimation \cite{bradley2003importance}. Finally, the results in Table \ref{tab:correlation_results} confirm that $\textbf{KF-}\bA(l)$ achieves the highest correlation between estimated and recorded signals. Notably, the $\textbf{LI-}\bA(l)$ algorithm performs particularly poorly in this measure. These results reinforce the efficacy of the proposed approach in estimating RIRs in real-world applications.
\section{Conclusions}
In this paper, we extend our previous method \cite{macwilliam2024state} for estimating the early part of RIRs in a time-variant acoustic scenario using a state-space model with the ISM incorporated into the state transition matrix. By segmenting longer trajectories into smaller linear segments, we address more complex scenarios while preserving the original assumptions. We validated both the original and extended methods with real measurements, demonstrating that they outperform both RIR interpolation and a purely data-driven state-space model using a transition factor. Additionally, we showed that our DTW-based parameter estimation for the derived state transition matrix improves the pairing and detection of reflection TOAs across RIRs.

\section{Acknowledgments}
This work was supported in part by 1) The European Research Council under the European Union's Horizon 2020 research and innovation program / ERC Consolidator Grant: SONORA (no. 773268). This paper reflects only the authors' views and the Union is not liable for any use that may be made of the contained information, 2) KU Leuven Internal Funds C14/21/075 ``A holistic approach to the design of integrated and distributed digital signal processing algorithms for audio and speech communication devices'', 3) The FWO Research Project: ``The Boundary Element Method as a State-Space Realization Problem'' (G0A0424N)
\bibliography{fa2025_template}

\end{document}